\documentclass[fleqn,twoside,amsfonts,amssymb]{article}
\usepackage{espcrc2}

\usepackage{graphicx}

\newcommand{\mibs}[1]{\mbox{\scriptsize{\bf #1}}}
\newcommand{\ave}[1]{{\langle#1\rangle}}

\newcommand{\citePRB}[3]{Phys.\ Rev.\ B {\bf #1} (#2) #3}
\newcommand{\citePRL}[3]{Phys.\ Rev.\ Lett. {\bf #1} (#2) #3}
\newcommand{\citeRMP}[3]{Rev.\ Mod.\ Phys. {\bf #1} (#2) #3}
\newcommand{\citeJPSJ}[3]{J.\ Phys.\ Soc.\ Jpn. {\bf #1} (#2) #3}

\newcommand{\citeEPJB}[3]{Eur.\ Phys.\ J. B {\bf #1} (#2) #3}
\newcommand{\citeJPF}[4]{J.\ Phys.\ (France) #1 {\bf #2} (#3) #4}
\newcommand{\citeIBID}[3]{ibid. {\bf #1} (#2) #3}

\newcommand{\veps}{\varepsilon}

\newcommand{\AmS}{{\protect\the\textfont2
  A\kern-.1667em\lower.5ex\hbox{M}\kern-.125emS}}

\title{Finite-Temperature Mott Transition in the Two-Dimensional Hubbard Model}

\author{S. Onoda\thanks{Corresponding author. Tel.: +81-3-5841-7904; fax.: +81-3-5841-7904.
{\it E-mail address:} sonoda@appi.t.u-tokyo.ac.jp}\address[MCSD]{Tokura Spin Superstructure Project, ERATO, Japan Science and Technology Corporation c/o,\\
        Department of Applied Physics, University of Tokyo, Tokyo 113-8656, Japan}%
        and
        M. Imada\address{Institute for Solid State Physics, University of Tokyo, Kashiwa, Chiba 277-8581, Japan}
       }
\begin{document}

\begin{abstract}
Mott transitions are studied in the two-dimensional Hubbard model by a non-perturbative theory of correlator projection that systematically includes spatial correlations into the dynamical mean-field theory. A nonzero second-neighbor transfer yields a first-order Mott transition at finite temperatures ending at a critical point.

\vspace{1pc}
\end{abstract}

\maketitle

The Mott transition 
 has been one of the fundamental issues in strongly correlated electrons \cite{RMP_Imada}. In spite of substantial progress \cite{RMP_Imada,DMFT}, its full understanding in low dimensions remains open.

Here, we propose a non-perturbative theoretical framework of correlator projection method (CPM) \cite{CPM}. The CPM combines the operator projection method \cite{OPM} with the dynamical mean-field approximation (DMFA) by restoring spatial correlations along the coarse-graining concept in the energy-momentum space with increasing the order of projection. It does not suffer from a cluster-size problem. Then, we obtain the Mott transition phase diagram of the two-dimensional (2D) Hubbard model in the parameter space of the Coulomb repulsion $U$, the second-neighbor transfer $t'$, the temperature $T$ (scaled by the nearest-neighbor transfer $t$) and the filling $\ave{n}$. A first-order Mott tansition surface with a critical end curve appears at half filling. Doping into the Mott insulator yields a diverging compressibility and/or a tendency to phase separation.

We consider the 2D Hubbard model. The formalism of CPM \cite{CPM} is as follows: The operator projection leads to series of Dyson equations,
\begin{eqnarray}
\Sigma_{n-1}(\omega,{\bf k})&=&G_{n-1}(\omega,{\bf k})\veps^{(n,n-1)}_{\mibs{k}}
\label{eq:2OP:Sigma_n-1}\\
G_{n-1}(\omega,{\bf k})&=&1/[\omega-\veps^{(n,n)}_{\mibs{k}}-\Sigma_n(\omega,{\bf k})]
\label{eq:2OP:G_n-1}
\end{eqnarray}
with $n=1$, $2,\cdots$.
$G_0=\Sigma_0$ is the electron Green's function and $\Sigma_1$ is the self-energy part with $\veps^{(1,0)}_{\mibs{k}}=1$.
Other coefficients up to the second order $n=2$ are given by
the Hartree-Fock dispersion $\veps^{(1,1)}_{\mibs{k}}=-t_{\mibs{k}}-\mu-U\ave{n}/2$, a source of Hubbard band splitting $\veps^{(2,1)}_{\mibs{k}}=U^2M$ with $M=\ave{n}(2-\ave{n})/4$, and $\veps^{(2,2)}_{\mibs{k}}=-\tilde{t}_{\mibs{k}}-\mu-U(1-\ave{n}/2)-\veps_{\rm cor}/M$. Here, $\mu$ is the chemical potential determined from the filling $\ave{n}$. $\veps_{\rm cor}=-\sum_{\mibs{x}',s}t_{\mibs{x},\mibs{x}'}\ave{(1/2-n_{\mibs{x},-s})(c^\dagger_{\mibs{x},s}c_{\mibs{x}',s}+h.c.)}/2$ is a local energy shift due to a correlated hopping process with $n_{\mibs{x},s}=c^\dagger_{\mibs{x},s}c_{\mibs{x},s}$ where $c_{\mibs{x},s}$ and $c^\dagger_{\mibs{x},s}$ are the electron creation and annihilation operators at a site ${\bf x}$ with a spin $s$, respectively. $t_{\mibs{k}}$ and $\tilde{t}_{\mibs{k}}$ are the Fourier transforms of the transfer $t_{\mibs{x},\mibs{x}'}$ and $\tilde{t}_{\mibs{x},\mibs{x}'}=t_{\mibs{x},\mibs{x}'}(\ave{n_{\mibs{x}}n_{\mibs{x}'}}/4+\ave{\vec{S}_{\mibs{x}}\cdot\vec{S}_{\mibs{x}'}}-\ave{\Delta^\dagger_{\mibs{x}}\Delta_{\mibs{x}'}})/M$, respectively, where $n_{\mibs{x}}$, $\vec{S}_{\mibs{x}}$ and $\Delta_{\mibs{x}}$ represent the charge, spin and local-pair operators, respectively. $\tilde{t}_{\mibs{k}}$ introduces ${\bf k}$ dependences of $\Sigma_1(\omega,{\bf k})$ mainly through the superexchange process and produces antiferromagnetic precursors. Here, we determine it from Ref. \cite{TPSC}.

To reproduce a Mott insulator in particle-hole asymmetric cases, the local dynamics of $\Sigma_2(\omega,{\bf k})$ must be obtained in a sufficiently correct manner. Then, instead of decoupling approximations \cite{OPM,MatsumotoMancini97}, we adopt the following generalized DMFA: (i) We calculate a local {\it normalized} self-energy part $G_{1,{\rm loc}}(\omega)=\frac{1}{N}\sum_{\mibs{k}}G_1(\omega,{\bf k})$ from an arbitrary $\Sigma_{2,{\rm loc}}(\omega)$ by replacing $\Sigma_2(\omega,{\bf k})$ with $\Sigma_{2,{\rm loc}}(\omega)$ in Eq. (2), (ii) calculate the {\it Weiss self-energy part} ${\cal S}_1(\omega)=\veps^{(2,1)}/[G_{1,{\rm loc}}^{-1}(\omega)+\Sigma_{2,{\rm loc}}(\omega)]$, and (iii) obtain the {\it Weiss Green's function} ${\cal G}_0(\omega,{\bf k})=1/[\omega-\veps^{(1,1)}_{\mibs{k}}-{\cal S}_1(\omega)]$. (iv) Within the iterative perturbation scheme, we calculate a new $\Sigma_{2,{\rm loc}}(\omega)$ as a Fourier transform of $\Sigma_{2,{\rm loc}}(\tau)=\frac{1}{N^3}\sum_{\mibs{k},\mibs{k}',\mibs{q}}\Gamma_{\mibs{k},\mibs{k}',\mibs{q}}{\cal G}_0(\tau,\mibs{k}+\mibs{q}){\cal G}_0(\tau,\mibs{k}'-\mibs{q}){\cal G}_0(-\tau,\mibs{k}')$ where $\Gamma_{\mibs{k},\mibs{k}',\mibs{q}}=4[t_{\mibs{k}'}-t_{\mibs{k}'-\mibs{q}}-t_{\mibs{k}+\mibs{q}}+\tilde{t}_{\mibs{k}}+4\veps_{\rm cor}/(\ave{n}(2-\ave{n}))]^2/(\ave{n}(2-\ave{n}))$. This loop continues to (i) until convergence is reached.

This second-order CPM reproduces a Mott insulator at low $T$'s for $t'=0$ at half filling and its single-particle spectra and momentum distributions show a remarkable similarity to numerical results \cite{Assaad}. As we introduce $t'\ne0$ with $U$ and $\ave{n}=1$ being fixed, the system undergoes a first-order Mott metal-insulator transition (MIT) at a low $T$ accompanied by a discontinuity in the double occupancy $\ave{n_\uparrow n_\downarrow}$. In the space of $(U,t',T)$, this first-order MIT boundary forms a surface \cite{CPM}. Within the surface with $U$ being fixed, the jump in $\ave{n_\uparrow n_\downarrow}$ decreases with increasing $T$ and disappears above a critical temperature $T_{\rm cr}$; the first-order surface ends at a critical curve at $T>0$. Then, without any spontaneous symmetry breaking as in two dimensions at $T>0$, the Mott transition shows a similarity to a liquid-gas phase transition, as in Refs. \cite{Castellani79,DMFT}. The MIT boundary at $T\to0$ agrees with numerical results \cite{KashimaImada01}.

\begin{figure}[tbh]
\includegraphics[width=7.2cm]{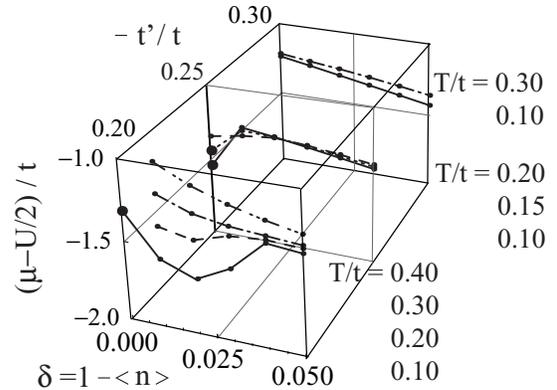}
\caption{$\mu$-$\delta$ curves obtained with $U/t=4$. The bold circles lie in the Mott-insulating side of the MIT surface at half filling.}
\label{fig:mu-n}
\end{figure}

Doping carriers into Mott insulators yields a metallic state characterized by an enhanced charge compressibility $\kappa=\partial\ave{n}/\partial\mu$. The present results of $\mu-\delta$ curves near half filling are shown in Fig. \ref{fig:mu-n} with $\delta=1-\ave{n}$. The slope $\partial \mu/\partial \ave{n}$ represents $1/\kappa$. Then, from high $T$'s, $\kappa$ increases with decreasing $T$ and shows a divergence towards the filling-controll MIT at low $T$'s. Even an instability to a phase separation appears as emergence of $\kappa<0$ at lower $T$'s. However, the issue if the phase separation is an artifact of the present approximation requires a further intensive study.

In summary, using the CPM that systematically includes spatial correlations into the DMFA, the MIT phase diagram of the 2D Hubbard model has been clarified in the $(U,|t'|,T)$ space with a first-order MIT surface with a $T>0$ critical end curve for $t'\ne0$ at half filling as well as a diverging charge compressibility and/or a tendency to a phase separation.

\end{document}